\documentstyle[aps,twocolumn,epsf]{revtex}

\newcommand{\beq}{\begin{equation}}
\newcommand{\eeq}{\end{equation}}
\newcommand{\ba}{\begin{eqnarray}}
\newcommand{\ea}{\end{eqnarray}}
\renewcommand{\a}{\alpha}

\newcommand{\m}{\mu}
\newcommand{\n}{\nu}

\begin{document}

\twocolumn[\hsize\textwidth\columnwidth\hsize\csname
@twocolumnfalse\endcsname

\vspace{-0.8cm}
\begin{flushleft}
\hspace{15cm}
DO-TH 98/17\\ \hspace{15cm}
THES-TP 98/7\\ \hspace{12cm}
%\today  \\
\end{flushleft}
\title{Finite-Size and Finite-Temperature Effects \\ in the Conformally
Invariant $O(N)$ Vector Model for $2<d<4$}

\author{{\bf{Anastasios C. Petkou}}$^{1,2}$ and
 {\bf{Nicholas D. Vlachos}}$^{2}$}
\address{$^{1}$ Institut f\" ur Physik, Universit\"at Dortmund, D-44221,
         Germany\\  $^{2}$ Department  of Theoretical Physics, Aristotle
         University of Thessaloniki, 54006 Thessaloniki, Greece}
\date{\today}
\maketitle

\begin{abstract}\noindent
We study the operator product expansion (OPE) of the auxiliary scalar
field $\lambda(x)$ with itself,
 in the conformally invariant $O(N)$ Vector Model
for $2<d<4$, to leading order in $1/N$ in a strip-like geometry with one
finite dimension of length $L$. We show that consistency of the
finite-geometry OPE with bulk OPE calculations requires the physical
conditions of, either finite-size scaling at criticality, or
finite-temperature phase transition.

\end{abstract} 

%\draft
%\vspace{-0.2cm}
%\pacs{PACS numbers: 03.75.F, 05.70.F, 64.60A.}
\vskip1.5pc]

%%%%%%%%%%%%%%%%%%%%%%%%%%%%%%%%%%%%%%%%%%%%%%%%%%%%%%%%%%%%%%%%%%%
%\section{General Theory}
%\label{sec1}
%%%%%%%%%%%%%%%%%%%%%%%%%%%%%%%%%%%%%%%%%%%%%%%%%%%%%%%%%%%%%%%%%%%

\section{Introduction}
Euclidean conformal field theory (CFT) describes systems at second
order phase transition points where the renormalised correlation
length $\xi$ becomes infinite \cite{Cardy1,Zinn-Justin}. For a theory
in the bulk, such points, if they exist, are reached when an
appropriate parameter (i.e. renormalised coupling) takes a
``critical'' value. The same theory defined in finite-geometry, i.e. a
hyper-strip with one finite dimension of length $L$, can also exhibit
critical behaviour when the renormalised coupling retains its bulk
critical value.  In such a case, the correlation length is not
infinite but it scales with $L$ and the system exhibits finite-size
scaling at criticality \cite{Cardy2}.  It may exist, however, another
possible ``critical state'' for the same theory in
finite-geometry. This is reached whenever it is possible to tune the
length scale $L$ (which is now interpreted as inverse temperature
$1/T$), to a ``critical'' value related to the renormalised coupling,
such that the renormalised correlation length diverges.  Such a case
would then describe a finite-temperature phase transition
\cite{Rosenstein}.

In the present note we quantify the above general remarks by studying
the conformally invariant $O(N)$ vector model in 
$2<d<4$, in
a hyper-strip geometry with one finite dimension of length $L$ and
periodic boundary conditions. Our tool for studying
finite-geometry critical systems is the
OPE which, by probing 
short distances is insensitive to the macroscopic size of the system.
In the case of the conformally invariant $O(N)$ vector model, the OPE
of the auxiliary field $\lambda(x)$ with itself in the
finite-geometry, should be consistent with the corresponding OPE
results in the bulk, which were 
studied in \cite{Ruhl,Tassos1}.  We show that, consistency of
the finite-geometry OPE with the OPE of the theory in the bulk
can be achieved in two different ways which correspond to two distinct
physical states of the system: {\bf{a)}} finite-size scaling at
criticality and {\bf{b)}} finite-temperature phase transition. Our
main result is a unified approach to finite-size scaling and
finite-temperature phase transitions based on CFT and OPEs. As a
by-product, we verify some old results of Cardy \cite{Cardy} on the
finite-size scaling of the free-energy density.
 
\section{The Conformally Invariant $O(N)$ Vector Model in
$2<d<4$} This model is one of the most extensively used testing grounds
for ideas related to CFT and phase transitions \cite{Ruhl,Kehrein,Halpern}.
Its ``effective'' partition function, obtained after integrating out
the fundamental $O(d)$-scalar and $O(N)$-vector fields $\phi^{\a}(x)$
with $\a =1,2,..,N$, reads \ba Z &=&\int ({\cal {D}}\sigma )\,\exp
\left[ -\frac{N}{2}S_{eff}(\sigma ,g)\right]\,, \label{eq1}\\
S_{eff}(\sigma ,g)&=&\mbox{Tr}[\ln (-\partial ^{2}+\sigma
)]-\frac{1}{g} \int {\rm d}^{d}x\,\sigma (x)\label{eq2}\,, \ea where,
$\sigma(x)$ is the auxiliary $O(d)$-scalar field and $g$ the coupling.
Setting $\sigma (x)=m^{2}+(i/\sqrt{N})\lambda (x)$, (\ref{eq1}) can be
calculated in a $1/N$ expansion provided that the gap equation
\begin{equation}
\frac{1}{g}=\int \frac{{\rm d}^{d}p}{(2\pi
)^{d}}\frac{1}{p^{2}+m^{2}}\,, \label{eq3}
\end{equation}
is satisfied. The leading-$N$ inverse propagator of $\lambda(x)$ is
\begin{equation}
\Pi ^{-1}(p^{2})=\frac{1}{2}\int \frac{{\rm d}^{d}q}{(2\pi
)^{d}}\frac{1}{(q^{2}+m^{2})[(q+p)^{2}+m^{2}]}\,.
\label{eq4}
\end{equation}
A non-trivial CFT, to any fixed order in $1/N$, is obtained for
$2<d<4$ by tuning the coupling to the critical value $1/g\equiv
1/g_{\ast }= (2\pi)^{-d}\int{\rm d}^{d}p/p^{2}$ \cite{Rosenstein}.
Then, the mass or inverse correlation length $m\equiv 1/ \xi=0$.

Putting the model in a hyper-strip geometry with one finite dimension
of length $L$ and periodic boundary conditions, does not affect its
renormalisability in the $1/N$ expansion
\cite{Zinn-Justin,Rosenstein}.  Explicit calculations are now more
involved since the momentum along the finite dimension takes the
discrete values $\omega_{n}=2\pi n/L$, $n=0,\pm 1,\pm 2,..$ and the
relevant integrals become infinite sums. For example, the gap equation
and inverse $\lambda(x)$ propagator read respectively
\begin{eqnarray}  \frac{1}{g}
& = &\frac{1}{L}\,\!\!\sum\limits_{n=-\infty }^{\infty }\int \frac{
{\rm d}^{d-1}p}{(2\pi
)^{d-1}}\frac{1}{p^{2}+\omega_{n}^{2}+M^{2}_{L}}\,, \label{eq5} \\ \Pi
_{L}^{-1}(p^{2},\omega_{n}^{2}) & = &
\frac{1}{2L}\,\!\!\sum_{m=-\infty }^{\infty }\int \frac{{\rm d}
^{d-1}q}{(2\pi )^{d-1}}
\left[\frac{1}{q^{2}+\omega_{m}^{2}+M^{2}_{L}}\right.  \nonumber \\ &
&
\hspace{0.6cm}\times\left.
\frac{1}{(p+q)^{2}+(\omega_{n}^{2}+\omega_{m}^{2})
+M^{2}_{L}}\right]\, ,\label{eq6} \ea where $M_{L}$ is a mass or
inverse correlation length parameter.

\section{The CFT Results}

Before studying the possible critical points in finite-geometry, we
recall the CFT results. The conformally invariant OPE of the auxiliary
field $\lambda(x)$ with itself for the theory in the bulk, has been
studied in the $1/N$ expansion in a number of articles
\cite{Ruhl,Tassos1}. Here, we confine ourselves to the leading order
in $1/N$ where the OPE reads 
\ba 
\lambda(x)\lambda(0) & = &
\frac{1}{x^{4}}+g_{\lambda} \frac{1}{ x^{2}}\lambda(0)\nonumber \\ & &
+\frac{d\,\Gamma(\frac{1}{2}d)}{
\pi^{\frac{1}{2}d}(d-1)C_{T}}\frac{x_{\mu }x_{\nu
}}{(x^{2})^{3-\frac{1}{2}d}} T_{\mu \nu }(0)+\cdots\, .\label{eq7} 
\ea
Only the most singular terms as $x\rightarrow 0$ are displayed on the
r.h.s. of (\ref{eq7}) which, in general, receives contributions from an
infinity of operators \cite{Ruhl,Kehrein}.  $C_{T}$ is the
normalisation coefficient
of the two-point function of the energy momentum tensor
$T_{\m\n}(x)$ and $g_{\lambda}$ is the coupling constant of the
three-point function $\langle\lambda\lambda\lambda\rangle$. Note that
to leading-$N$ the dimension of $\lambda(x)$ is 2.

As mentioned in the introduction, the finite-geometry OPE of
$\lambda(x)$ with itself is expected to retain the bulk form
(\ref{eq7}), when the system is at a conformally invariant point. Then,
finite-geometry corrections to the bulk form $1/x^{4}$ of the
$\lambda(x)$ propagator are obtained from (\ref{eq7}), through
non-vanishing expectation values of the operators in the r.h.s. In
general, these expectation values are unknown. However, using only the
conformal invariance of the critical theory, Cardy \cite{Cardy} showed
that there exists a relation between the expectation value of the
diagonal elements of $T_{\m\n}(x)$ and the finite-size scaling
correction to the
free-energy density $f=-\ln Z/V$ of the system at criticality. Namely,
\beq
f_{\infty}-f_{L}=
\tilde{c}\frac{\Gamma(\frac{1}{2}d)\zeta(d)}{\pi^{\frac{1}{2}d}L^{d}}
=\frac{1}{d-1}\langle T_{11}\rangle =-\langle T_{ii}\rangle\,
.\label{eq8} \eeq 
Here, $\tilde{c}$ is a parameter which characterises the specific CFT
model under consideration. It has been proposed
\cite{Cardona,Fradkin} to be a possible generalisation of the central
charge in $d>2$. 

Upon transforming the expectation value of (\ref{eq7}) to momentum
space \cite{note} and inverting, we obtain by virtue of (\ref{eq8})
the inverse propagator of $\lambda(x)$ in finite geometry as
\begin{eqnarray}
\Pi_{L}^{-1}(p^{2},\omega _{n}^{2}) &=&\frac{K_{d}}{ (p^{2}+\omega
_{n}^{2})^{2-\frac{1}{2}d}}\left\{1-\frac{4\left( {\textstyle{
\frac{1}{2}d-2}}\right)g_{\lambda}}{p^{2}+\omega _{n}^{2}}\langle
{\lambda} \rangle \right.  \nonumber \\ &&\hspace{-1.5cm}
\left.-\frac{2^{d+2}\Gamma (d-1)}{\Gamma(\frac{1}{2}d)\,\Gamma
(2-\frac{1}{2}d)}\,\frac{\zeta (d)\tilde{c}}{L^{d}\,N}\, \frac{
C_{2}^{\scriptstyle{\frac{1}{2}d-1}}(y)}{(p^{2}+\omega
_{n}^{2})^{\frac{1}{2}d}}+....\right\},\label{eq9}
\end{eqnarray}
where, $n=0,\pm 1,\pm 2,..$,
$y=\omega_{n}/\sqrt{p^{2}+\omega_{n}^{2}}$, $C_{\n}^{\m}(y)$ is the
Gegenbauer polynomial and the leading-$N$ relation \cite{Tassos1}
$C_{T}/N=d\,\Gamma^{2}(\frac{1}{2}d)/4\pi^{d}(d-1)$ was used. $K_{d}$
is a $d$-dependent constant whose value is not important here.

Our next task is to evaluate (\ref{eq6}) and cast it in a form
suitable for comparison with (\ref{eq9}). Also, care must be taken so
that $M_{L}$ satisfies the gap equation (\ref{eq5}). The calculational
details are given in \cite{Tassos2}. Our results show that there are
two ways for obtaining a large-momentum expansion of (\ref{eq6})
consistent with (\ref{eq9}), which correspond to two distinct
physical states of the system.

\section{Finite-Size Scaling}
One way to obtain a critical theory in finite-geometry is by tuning the
coupling $1/g$ in (\ref{eq5}) to its bulk critical value
$1/g_{*}$. Then, for the whole range $2<d<4$, the (renormalised) gap
equation (\ref{eq5}) has a finite solution which gives the finite-size
scaling of the inverse correlation length $M_{L}\equiv M_{*}\sim 1/L$
as shown in Fig.1 (see also \cite{Chamati}).

\begin{figure}[h]
\epsfxsize=6cm \epsffile{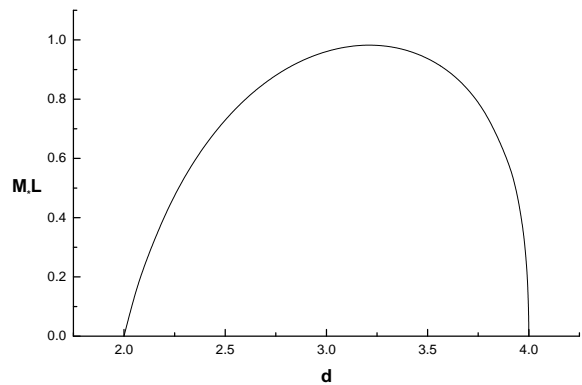}
\caption{The finite-size scaling of $M_{*}$}
\end{figure}

Expression (\ref{eq6}) can be evaluated for a general $M_{L}$, but for
$M_{L}=M_{*}$ it takes the simple form
\begin{eqnarray}
\Pi_{L}^{-1}(p^{2},\omega_{n}^{2}) & = &\frac{K_{d}}{(p^{2}+\omega
_{n}^{2})^{2-\frac{1}{2}d}}\Bigg\{ 1+{\textstyle{2(d-3)}}
\frac{M_{*}^{2}}{p^{2}+\omega _{n}^{2}} \nonumber \\ & &
\hspace{-2.2cm}  +\frac{2^{2d}\Gamma (\frac{1}{2}d-\frac{1}{2})
C_{2}^{\frac{1}{2}d-1}(y)M_{*}^{d}}{\sqrt{\pi }\Gamma (d)\Gamma
(2-\frac{1}{2}d)(p^{2}+\omega
_{n}^{2})^{\frac{1}{2}d}}{\cal{F}}_{d}(LM_{*}) +.. \Bigg\}\,,\label{eq10}
\\ {\cal{F}}_{d}(LM_{*}) & = & \frac{1-d}{d}{\cal {I}} _{0}-{\cal
{I}}_{1}\,, \label{eq11} \\ {\cal I}_{n} & = & \int_{1}^{\infty }{\rm
d}t\frac{(t^{2}-1)^{\frac{1}{2}\left( d-3\right) +n}}{e^{LM_{*
}t}-1}\quad \,.\label{eq12}
\end{eqnarray}

Consistency of (\ref{eq10}) with (\ref{eq9}) requires \ba \langle
\lambda\rangle & = &
-\frac{M_{*}^{2}(d-3)}{2g_{\lambda}(\frac{1}{2}d-2)}\,,\label{eq13}\\
f_{\infty}-f_{L} & = &N\frac{M_{*}^{d}
\Gamma(\frac{1}{2}d)}{\pi^{\frac{1}{2}d}\Gamma(d)} \,\left(
{{\frac{d-1 }{d}}}{\cal {I}}_{0}+{\cal {I}}_{1}\right) \,.\label{eq14}
\ea To obtain (\ref{eq14}) we have used (\ref{eq8}).  One can show
\cite{Tassos2} that (\ref{eq13}) is consistent with known CFT
calculations of the coupling constant $g_{\lambda}$ and the OPE of the
fundamental $O(N)$-vector field $\phi^{\a}(x)$ with itself
\cite{Ruhl,Tassos1}. On the other hand, (\ref{eq14}) is exactly what
it is obtained from a direct calculation of the finite-size scaling
correction to
the free-energy density at criticality \cite{Sachdev}. In particular,
for $d=3$ (\ref{eq14}) reproduces the intriguing result
$\tilde{c}/N=4/5$ \cite{Sachdev,Rebhan}, while for $2<d<4$ a plot of
$\tilde{c}/N$ is depicted in Fig.2.

\begin{figure}[h]
\epsfxsize=6cm 
\epsffile{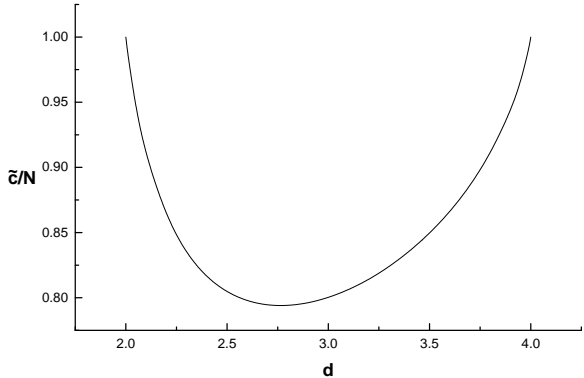}
\caption{The parameter $\tilde{c}/N$ for general $2<d<4$.}
\end{figure}

It is important to mention that the relatively simple form of
(\ref{eq10}) follows from a non-trivial cancellation that takes place
in the large momentum expansion of (\ref{eq6}) for $M_{L}=M_{*}$ 
\cite{Tassos2}. It can
be shown that this cancellation is related to the so-called ``shadow
singularities'' cancellations \cite{Ferrara,Ruhl,Tassos1} and also to
the gap equation (\ref{eq5}).

\section{Finite-Temperature Transition}
The second way to obtain a finite-geometry critical theory is by
tuning the renormalised coupling to a certain value such that the
inverse correlation length $M_{L}\equiv M_{*}=0$. In the present case,
this can only be achieved for $3<d<4$, since for $2<d\leq 3$ the gap
equation (\ref{eq5}) is divergent when $M_{L}=0$ \cite{Rosenstein},
implying that the massless phase of the theory is saturated.  In other
words, there is no finite critical temperature for the
$O(N)\rightarrow O(N-1)$ phase transition, in agreement with the
Mermin-Wagner-Coleman theorem \cite{Zinn-Justin,Sachdev}.

When $3<d<4$, the (renormalised) gap equation (\ref{eq5}) is satisfied
for $M_{L}\equiv M_{*}=0$ if the renormalised coupling \beq
\frac{1}{g}=\frac{\Gamma(1-\frac{1}{2}d)}{(4\pi)^{\frac{1}{2}d}}m_{*}^{d-2}
\,,\label{eq15} \eeq is such that \beq \frac{1}{L_{*}^{d-2}}=
T_{cr}^{d-2}=\frac{\Gamma(1-\frac{1}{2}d)}{2^{d-1}\Gamma(\frac{1}{2}d
-1)\zeta(d-2)} m_{*}^{d-2}\,,\label{eq16} \eeq for some ``critical
temperature'' $T_{cr}$.  The mass parameter $m_{*}$ is related to the
mass of the fundamental fields $\phi^{\a}(x)$ of the bulk theory
\cite{Kapusta}.  This agrees with \cite{Svaiter} for $3<d<4$ and
reproduces the leading-$N$ critical temperature for the $O(N)$
$\phi^{4}$ theory in $d=4$ up to an appropriate normalisation of the
$\phi^{4}$ coupling. Hence, the critical theory now corresponds to a
finite-temperature phase transition.

Furthermore, the critical theory is also conformally invariant since
for $M_{L}\equiv M_{*}=0$, the inverse propagator (\ref{eq6}) has the
large momentum expansion
\begin{eqnarray}
\Pi_{L}^{-1}(p^{2},\omega_{n}^{2})& = &\frac{K_{d}}{(p^{2}+\omega
_{n}^{2})^{2-\frac{1}{2}d}}\Biggl\{ 1+\nonumber \nonumber \\ & &
\hspace{-1.5cm}+ \frac{2^{2d-3}\,\Gamma
(\frac{1}{2}d-\frac{1}{2})}{\sqrt{\pi }\,\Gamma (2-
\frac{1}{2}d)}\frac{\zeta (d-2)}{[L_{*}^{2}(p^{2}+\omega
_{n}^{2})]^{\frac{1}{2} d-1}} \nonumber \\
&&\hspace{-2cm}-\frac{2^{2d}\Gamma (\frac{1}{2}d-\frac{1}{2})}{\sqrt{
\pi }\Gamma (2-\frac{1}{2}d)}\frac{\zeta (d)}{[L_{*}^{2}(p^{2}+\omega
_{n}^{2})]^{\frac{1}{2}d}}C_{2}^{\frac{1}{2}d-1}(y)+..
\Biggl\}\,.\label{eq17}
\end{eqnarray}
Indeed, (\ref{eq17}) is similar to the conformal OPE (\ref{eq9}).
Consistency then of (\ref{eq17}) and (\ref{eq9})
requires $\tilde{c}/N=1$.  This value of $\tilde{c}$ corresponds to a
free CFT of $N$ massless scalars. However, the critical theory under
study is not trivial since it is expected that $\tilde{c}$ receives
$1/N$ corrections which will alter the above leading-$N$ value.

An important difference between (\ref{eq17}) and (\ref{eq9}) is that
the field $\lambda(x)$ does not show up in (\ref{eq17}) - instead the
``shadow field'' of $\lambda(x)$ appears, which has leading-$N$
dimension $d-2$. To this end, it is worth pointing out that scalar
representations of the conformal 
algebra with dimensions $\eta$ and $d-\eta$ are equivalent in $d>2$
\cite{Ferrara}.

\section{Results and Outlook}
In this note, we investigated the intimate relationship connecting
finite-size scaling at criticality and conformal OPEs and we presented
a method for calculating the finite-size scaling correction to the 
free-energy density at criticality from two-point functions.  Also, we
discussed a critical 
point of the $O(N)$ vector model in $3<d<4$, where the renormalised
mass or inverse correlation length is zero and argued that it
corresponds to the physical situation of a finite-temperature phase
transition. It may be interesting to point out that the physical
situation of finite-temperature phase transition corresponds
effectively to a
CFT in $d-1$ dimensions, since critical
fluctuations with infinite correlation length do not ``see'' the
finite dimension.

Possible extensions of our work would be the study of other
conformally invariant models in finite geometries for $2<d<4$, such as
the Gross-Neveu or $CP^{N-1}$ models \cite{Gracey}. One could also
consider other finite geometries i.e. with two finite dimensions
\cite{Cardy4}, where
there is evidence \cite{Henkel} that there exist interesting similarities with
two-dimensional CFT, at least for antiperiodic boundary conditions.

\acknowledgments

\small 
A.C.P. would like to thank the organizers of ``TFT 98'' in
Regensburg for the stimulating atmosphere and for giving him the
opportunity to present this work. He would also like to acknowledge
very interesting discussions with A. Rebhan and G. Semenoff. This work
was partially supported by PENED/95 K.A. 1795 research grant and DAAD
A/98/19026 research fellowship.


\normalsize

\begin{references}
\bibitem{Cardy1} J. L. Cardy in ``Champs, Cordes et Ph\' enom\` enes
Critiques'', (E. Br\' ezin and J. Zinn-Justin, Eds), North Holland,
Amsterdam, 1989.
\bibitem{Zinn-Justin} J. Zinn-Justin,`` Quantum Field Theory and
Critical Phenomena'', 2nd ed. Clarendon, Oxford, 1993.
\bibitem{Cardy2} J. L. Cardy Ed. ``Finite-Size Scaling'', North
Holland, Amsterdam, 1988.
\bibitem{Rosenstein}
B. Rosenstein, B. J. Warr and S. H. Park, Nucl. Phys. B 336 (1990) 435.
\bibitem{Ruhl}
K. Lang and W. R\" uhl, Nucl. Phys. B 402 (1993) 573; Z. Phys. C 61
(1994) 495.
\bibitem{Tassos1}
A. C. Petkou Ann. Phys. 249 (1996) 180.
\bibitem{Kehrein}
S. K. Kehrein and F. Wegner, Nucl. Phys. B 424 (1994) 521.
\bibitem{Halpern}
M. B. Halpern, Phys. Lett. B 137 (1984) 382.
\bibitem{Cardy} 
J. L. Cardy, Nucl. Phys. B 290 (1987), 355.
\bibitem{Cardona}
X. Vilasis-Cardona, Nucl. Phys. B 435 (1995) 735.
\bibitem{Fradkin}
A. H. Castro-Neto and E. Fradkin, Nucl. Phys. B 290 (1993) 525,\\
M. Zabzine, hep-th/9705015,\\
D. M. Danchev and N. S. Tonchev, cond-mat/9806190. 
\bibitem{note}
Althought one could follow an $x$-space approach to OPE studies, we
prefer to work in momentum space as it is customary for
finite-temperature and finite-size scaling studies (see
e.g. \cite{Sachdev,Kapusta}).
\bibitem{Tassos2}
A. C. Petkou and N. D. Vlachos, hep-th/9803149
\bibitem{Chamati}
H. Chamati, E. S. Pisanova and N. S. Tonchev, Phys. Rev. B 57 (1998) 5798.
\bibitem{Sachdev}
S. Sachdev, Phys. Lett. B 309 (1993) 285.
\bibitem{Rebhan}
I. T. Drummond, R. R. Horgan, P. V. Landshoff and A. Rebhan,
Nucl. Phys. B 524 (1998) 579.
\bibitem{Ferrara}
S. Ferrara and G. Parisi, Nucl. Phys. B 42 (1972) 281.
\bibitem{Kapusta}
A. Bochkarev and J. Kapusta, Phys. Rev. D 54 (1996) 4066.
\bibitem{Svaiter}
G. N. J. A\~ na\~ nos, A. P. C. Malbouisson and N. F.
Svaiter, hep-th/9806027.
\bibitem{Gracey}
J. A. Gracey, Int. J. Mod. Phys. A 6 (91) 395; A 9 (94) 576, 727;
Phys. Lett. B 279 (92) 292; B 308 (93) 65; Z. Phys. C 59 (93) 243.
\bibitem{Cardy4}
J. L. Cardy, J. Phys. A 18 (85) L757.
\bibitem{Henkel}
M. Henkel, ``Conformal Invariance and Critical Phenomena'',
Springer-Verlag, Heidelberg, to appear.\\
M. Weigel and W. Janke, cond-mat/9809253.

\end{references}
\end{document}